\documentclass[twocolumn,aps]{revtex4}
\usepackage{graphicx}

\begin{document}
\title{No-clicking Event in the Quantum Key Distribution}

\author{Won-Young Hwang,$^*$
Intaek Lim, and Jongwon Park}

\affiliation{ Department of Physics Education, Chonnam National
University, Kwangjoo 500-757, Republic of Korea}

\begin{abstract}

We discuss the `no-clicking' event, which is harmful for legitimate
users in Bennett-Brassard 1984 quantum key distribution: We describe
an attack where no-clicking events are utilized in the same way as
double-clicking events are utilized in the quantum Trojan-pony
attack. We discuss how to deal with the no-clicking events. We
discuss how to estimate the security of the protocol against the
proposed attack by using a formula involved with the fraction of
adversarial removals of events.

\end{abstract}
\maketitle
\section{Introduction}\label{sec:intro}

Information processing with quantum systems, e.g., quantum
cryptography \cite{Wie83,Ben84,Joo05}, quantum computation \cite
{Sho94}, and quantum metrologies \cite{Yue86,Dow98}, enables us to
do some tasks that we cannot do with its classical counterparts. In
addition to the practical importance, this fact has big theoretical
implications.

The Bennett-Brassard 1984 (BB84) quantum key distribution (QKD)
protocol \cite{Ben84} is one of the most promising quantum
information processing techonology. Security of the BB84 protocol
was studied in the ideal case where the source, the channel, and the
detector are all perfect \cite{Yao95}. However, no actual device is
perfect. Later, security of the BB84 protocol in the case where the
channel was imperfect while the source and the detector were perfect
was given \cite{May00,Bih00,Sho00,Ham03,Hwa05}.

An imperfect source gives rise to the problem of
photon-number-splitting (PNS) attack \cite{Hut95,Lut00,Bra00}.
Methods to deal with the problem are given in Refs. \cite{Ina01} and
\cite{Got04}. However, the problem of the PNS attack becomes more
serious when the imperfect source is combined with loss
\cite{Hut95,Lut00,Bra00}. Thus, a practicable long-distance BB84 QKD
system, which usually has high loss, is not secure against the PNS
attack. The decoy method \cite{Hwa03,Wan04,Lo04} and the SARG04
protocol \cite{Sca04} are two independent ways to overcome the
problem of the PNS attack combined with high loss. The long-distance
BB84 protocol supplemented with a decoy can be secure as long as
detectors are perfect.

A remaining step is, therefore, to resolve problems due to imperfect
detectors. Imperfect detectors give rise to undesired events, e.g.
double-clicking and no-clicking. A no-clicking event can be easily
overlooked because they are not exposed. However, no-clicking events
can give rise to a post-selection effect as in the case of
detection-loopholes \cite{Sel88} for Bell's inequality violation.

The purpose of this paper is to discuss problems due to no-clicking
in the BB84 protocol. Our presentation is in the following order: In
Section II, we describe why the no-clicking event is problematic. In
Section III, we discuss how to deal with the no-clicking event. In
Section IV, we discuss how to estimate the security of the protocol.
In Section V, we conclude.

\section{Problematic no-clicking event}\label{sec:NDE}

Let us define terminology. A {\it loss-event} is the case where no
detector of Bob clicks even when a quantum carrier with a bit of
key, e.g., a pulse of photons, was sent by Alice. Here, Alice and
Bob are the legitimate sender and receiver of quantum carriers,
respectively. The loss-event may be either due to a {\it
channel-loss} event in which the quantum carrier was lost in the
channel or due to a {\it no-clicking} event in which the detector
did not click for certain reasons, e.g., it failed to click because
of imperfection. That is, either channel-loss or no-clicking can
give rise to a loss-event. However, channel-loss and no-clicking
events cannot be discriminated by Alice and Bob.

The possibility that no-clicking event was problematic was first
noted in Ref. \cite{Bra00}. In order to deal with the no-clicking
event, they "conservatively" assume that "Eve has control ... on
$\eta_B$ (detection efficiency),..." (Here Eve denotes an
eavesdropper.) What they mean is the following: As mentioned above,
a loss event may be due to either a channel-loss event or a
no-clicking event. What they assume is that detectors are perfect;
thus, all loss events are due to channel losses. Here channel losses
are due to Eve's action.

However, imperfect detectors give rise to post-selection effects
that may be harmful for Alice and Bob. This is implied by a few
papers. The post-selection effect by imperfect detectors is
mentioned in Section XIII of Ref. \cite{Got04}. Other interesting
post-selection effects are discussed in Ref. \cite{Cur04} and Refs.
\cite{Mak06} and \cite{Qi07}. The Referencs \cite{Mak06} and
\cite{Qi07} overlap with our paper, but are different, as will be
discussed in Section V.
How to coherently deal with all post-selection effects due to the
no-clicking event, including those works
\cite{Got04,Cur04,Mak06,Qi07}, is an open problem. In Ref.
\cite{Got04}, they discuss the problem with detector efficiency.
They give a formula (Eq. (58) of Ref. \cite{Got04}) for the key
generation rate. However, a fraction '$f$' of random removal is not
provided. Now let us see at how a no-clicking event can be
problematic in more detail.

First, let us see why Fred, a friend of Eve in detectors, is
introduced \cite{Got04}. Imperfectness in detector is not something
that is supposed to be in full control of Alice and Bob. It can be
that the imperfectness gives rise to systematic errors that happen
to be useful for Eve. Thus, Alice and Bob may consider the worst
case in which a hypothetical being, Fred, who wants to help his
friend Eve has full control of the detector. He can turn the
detector on or off as he wants. Because it is not excluded that Fred
can get information about the basis of the detector, he is supposed
to know the basis. One might say that if a friend of Eve has access
to Bob's detector, why doesn't the friend read the final plain text
of Bob directly and sends it to Eve \cite{Aci04}. That powerful
friend can break the whole system of Alice and Bob, of course.
However, the Fred we consider is less powerful than this friend of
Eve. Fred is confined in a detector, so he cannot observe something
outside the detector by any means except by quantum carriers given
to him via a window of the detector.

An alliance of Eve and Fred can cheat a careless Alice and Bob, for
even moderate loss, who take no account of no-clicking events: Like
in the case of the quantum `Trojan pony' attack \cite{Got04}, Eve
performs an opaque (intercept-resend) attack \cite{Gis02} in the
following way: Eve does not know the basis of Alice, of course.
However, she does a measurement that is randomly picked up between
$Z$ and $X$ for each photon pulse sent by Alice. Here, $Z$ and $X$
are measurements in the $z$ basis, $\{|0\rangle, |1\rangle \}$ and
in the $x$ basis, $\{|\bar{0}\rangle, |\bar{1}\rangle \}$,
respectively.
Here, $|0\rangle$ and $|1\rangle$ are two orthogonal states;
$\langle0|1\rangle=0$, $|\bar{0}\rangle= (1/\sqrt{2})(|0\rangle+
|1\rangle)$, and $ |\bar{1}\rangle= (1/\sqrt{2})(|0\rangle-
|1\rangle)$. When the outcome of the $Z$ measurement is $0$ ($1$),
Eve prepares multiple copies of $|0\rangle$ ($|1\rangle$), namely
$|0\rangle^{\otimes N}$ ($|1\rangle^{\otimes N}$). Then, she
forwards them to Bob. In the same way, when the outcome of the $X$
measurement is $0$ ($1$), Eve prepares multiple copies of
$|\bar{0}\rangle$ ($|\bar{1}\rangle$), namely
$|\bar{0}\rangle^{\otimes N}$ ($|\bar{1}\rangle^{\otimes N}$). Then,
she forwards them to Bob. The strategy of Fred is the following: He
splits the state $|k\rangle ^{\otimes N}$ ($k= 0, 1, \bar{0},
\bar{1}$) to $|k\rangle ^{\otimes N-1}$ and $|k\rangle$. When $N-1$
is large enough, Fred can identify the state with high reliability.
If the basis of the state that Eve has forwarded is the same as that
of the detector, Fred gives the remaining state, $|k\rangle$, to the
detector. If different, Fred turns off the switch of the detector.
Now let us see how Eve can eavesdrop without being detected. Note
that the only case in which Eve's action can be caught is when the
basis of Eve and that of Alice and Bob alliance do not match; Alice
and Bob happen to adopt the same basis and Eve happens to adopt the
other basis. However, in the above attack, whenever a non-matching
case happens Fred turns off the detector. Therefore, Eve's action is
not detected at all even if she performs the above attack for every
quantum carrier if Alice and Bob simply discard no-clicking events.

The above attack using no-clicking event is in parallel with the
quantum Trojan pony attack \cite{Got04}. In the quantum Trojan pony
attack, Eve attempts to nullify non-matching cases by making Bob's
detector double-click. If Alice and Bob simply discard the data for
a double-click event, the quantum Trojan pony attack reduces to an
attack using no-clicking event in which Alice and Bob discard the
data for a no-clicking event. However, a double-clicking event can
be detected directly. Alice and Bob take into account a
double-clicking event in their estimate of the security
\cite{Got04,Lut00}. A no-clicking event can be detected indirectly:
When the detector does not click when it is supposed to, we say that
a no-clicking event is detected. We term the attack using a
no-clicking event as a quantum `Trojan-dark-pony' attack. An easy
way to maintain security is for Alice and Bob to regard a
no-clicking event as an `error event' that contributes to the
quantum bit error rate (QBER) \cite{Gis02}. In this case, however,
loss lower-bounds the QBER. Thus, even for a moderate loss, there is
no secure protocol even if everything else is perfect. Thus, this
method is not practical.

\section{How to deal with no-clicking events}\label{sec:NDE}


Before we discuss how to maintain security, we have to characterize
detectors for a no-clicking event. However, in order to characterize
a detector, we need to find a principle governing the behavior of
the detector. At first look, one might think that it is difficult to
find the principle governing behavior of detectors for a no-clicking
event because a no-clicking event is neither a desired nor an ideal
process. However, this is not the case.

 {\it Proposition $1$:
a no-clicking event can be dealt with by using a normal quantum
measurement, POM. That is, a certain positive-operator is assigned
for a no-clicking event.}

Proposition$1$ is already adopted in Ref. \cite{Cur04}. However, let
us describe why Proposition 1 is valid, for those who are not
convinced. Usually, a quantum detector is a `black-box' that gives
rise to macroscopically distinct events for microscopic inputs. For
example, a photon polarization detector gives clicking of a photon
detector between two photon detectors as an output for an input
photon. However, although a no-clicking event literally does not
give a noticeable event, a no-clicking event can still be
macroscopically distinguished from other events. Thus, there is no
difficulty in assigning a number, say $0$, to a no-clicking event,
like in the case of other noticeable events. With Proposition $1$,
we can characterize a no-clicking event. What we have to do is to
identify the positive-operator corresponding to a no-clicking event
by repeated measurements.

Let us now discuss how to maintain security for a no-clicking event.
First let us introduce a quantity $\Delta$, the fraction of
adversarially removed events, that is important in estimating the
security of a protocol. In Ref. \cite{Got04}, they consider a
hypothetical situation where Fred can freely remove some cases that
would lead to the QBER, for example, as he does in non-matching
cases in the quantum Trojan-dark-pony attack above. If Alice and Bob
have no knowledge on how many times Fred did the adversarial
removal, then they have to assume the worst case that Fred did the
removal all of the instances; thus, clearly the protocol is not
secure. However, if Alice and Bob know a bound on the fraction of
adversarially removed event, $\Delta$, and $\Delta$ is small enough,
then the security of the protocol can be recovered with a reduced
key generation rate depending on the fraction $\Delta$ \cite{Got04}:
The larger the fraction $\Delta$ is, the smaller the key generation
rate is. Therefore, the problem of estimating security reduces to
how to estimate the fraction $\Delta$.

However, although Eve does not know the identity of the quantum
carriers that Alice has sent, it is Eve's freedom that she replace
Alice's quantum carriers by any other quantum carriers. In the
quantum Trojan-dark-pony attack, for example, it can be that a
quantum carrier $|\bar{0}\rangle$ is replaced by either
$|0\rangle^{\otimes N}$ or $|1\rangle^{\otimes N}$ if the $Z$
measurement is chosen. Therefore, we must analyze Bob's detector for
all different states of $N$ quantum bits. If the number of quantum
bits $N$ is unlimited, the analysis is impossible. Thus, we assume
that the number of quantum bits $N$ is bounded by a certain number
$M$ so that the analysis can be done. This assumption amounts to the
assumption that there is a bound on light intensity at Bob's site,
which can be checked by Bob. With the assumption, what we have to do
is to analyze all states of quantum bits whose number is less than
$M$. Then, by repeated measurements, we estimate the
positive-operator corresponding to a no-clicking event of Bob's
detector in each basis $b$, where $b= z, x$. With the
positive-operator thus obtained, we can calculate the detector
efficiency, $\eta_b(|\psi\rangle)$, for each state $|\psi\rangle$
and basis $b$.

Let us consider the simplest case where $\eta_b(|\psi\rangle)= 1$
for all states $|\psi\rangle$ and bases $b$. In this case, Fred has
no chance to turn off the switch; thus, the fraction $\Delta$ is
zero. Next, let us consider a case where $\eta_b(|\psi\rangle)=
\eta_0$ for all states $|\psi\rangle$ and bases $b$. Here, $\eta_0$
is a real number between $0$ and $1$. Here, the question is how
large the fraction $\Delta$ is. In this case, it is clear that the
fraction $\Delta$ is bounded by $1-\eta_0$. This corresponds to the
simple, but impractical, method above where Alice and Bob regard a
no-clicking event as an 'error event' contributing to the QBER.
Thus, we need to get a tighter bound for the fraction $\Delta$. It
appears that it is hard to do so because there is no way to
discriminate accidental removals by imperfection from intentional
removals by Fred. However, this is not the case as far as the
quantum Trojan-dark-pony attack is concerned.
This can be explained as follows: Let us recall that the basic
strategy of the quantum Trojan-dark-pony attack is that Fred
suppresses (enhances) what is advantageous (disadvantageous) for
Alice and Bob. More specifically, Fred turns off the switch of Bob's
detector in non-matching cases because only non-matching cases
contribute to the QBER, and he does nothing in matching cases
because in matching cases, Eve can obtain information on the key of
Alice and Bob without contributing to the QBER at all. However, in
this case where all $\eta_b(|\psi\rangle)$'s are identically
$\eta_0$, it is not that all no-clicking events are adversarial
removals. That is, some removals are rather {\it friendly} for Alice
and Bob: Each state is removed with a fixed rate $1-\eta_0$,
regardless of basis $b$. For example, even when $|\psi\rangle =
|1\rangle^{\otimes n}$ ($n < M$) is the input state for Bob's
detector as in the quantum Trojan-dark-pony attack, the removal
rates for two bases are both $1-\eta_0$ because
$\eta(|1\rangle^{\otimes n})_z=\eta(|1\rangle^{\otimes n})_x=
\eta_0$. Here, removals when the detector is in the $x$ basis are
adversarial for Alice and Bob, but those when the detector is in $z$
basis are friendly for Alice and Bob because this is the matching
case advantageous for Eve. The effect of friendly removals is
opposite that of adversarial removals, and here the numbers of
friendly and adversarial removals are the same. Therefore,
effectively, the fraction $\Delta(|1\rangle^{\otimes n})$ for the
state $|1\rangle^{\otimes n}$ is bounded by
$|\eta(|1\rangle^{\otimes n})_z-\eta(|1\rangle^{\otimes n})_x|=0$
regardless of how those events are removed, accidentally or
intentionally. However, for other states, for example,
$(|0\rangle+|\bar{0}\rangle)^{\otimes n}$, whether an attack using
the state is adversarial or friendly is not as clear as in the above
case. A removal of a certain state for a basis is partially
adversarial while the removal of a state for the other basis is
partially friendly, and similar arguments apply.

Let us now consider the general case where $\eta_b(|\psi\rangle)$
depends on the state $|\psi\rangle$. For each state $|\psi\rangle$,
we have a relation that
\begin{equation}
\label{A} \Delta(|\psi\rangle) \leq
|\eta_z(|\psi\rangle)-\eta_x(|\psi\rangle)|,
\end{equation}
as shown above. Then the bound $\Delta$ for all states is simply the
maximal one among all $\Delta(|\psi\rangle)$'s,
\begin{equation}
\label{B} \Delta \leq \mbox{Max} \{\Delta(|\psi\rangle)\}.
\end{equation}
Our recipe in Eq. (\ref{B}) is still quite loose bound for $\Delta$.
One reason is that it is not that a state that gives maximal
$\Delta$ is the only one utilized in the actual attack. In other
words, the situation can be that the state that gives maximal
$\Delta$ is not so effective in the eavesdropping. However, Eq.
(\ref{B}) is much tighter than the simple method above where all
no-clicking events are treated as error events. It might be that for
some detectors, the bound in Eq. (\ref{B}) is too large for a
protocol to be secure. However, it seems to be possible to design
detectors such that the no-clicking rate of each state does not
depend on the basis, in which case the fraction $\Delta$ is zero. In
other words, if loss of each state is independent of the basis then
the fraction $\Delta$ is zero. It is notable, however, that even if
the {\it overall} loss is independent of the basis, it might be that
the the fraction $\Delta$ is not zero because it is an absolute
value of $\eta_z(|\psi\rangle)-\eta_x(|\psi\rangle)$ that bounds the
fraction $\Delta$ in Eq. (\ref{A}).

The bound in Eq. (\ref{B}) can be easily translated to a bound for
the quantum Trojan-pony attack by only replacing the detector
efficiency $\eta$ with the probability of a non-double-clicking
event. In this case, however, the bound in Eq. (\ref{B}) is not that
useful because it is not easy to make the bound small in most
natural designs of detectors.

\section{Discussion and Conclusion}\label{sec:NDE}
The results in Ref. \cite{Mak06} overlap with ours. Let us see how
they differ. What is the same is that both escape Alice and Bob's
test by keeping the detector from clicking in the case of a
non-matching basis. However, what we consider is more general that
what they considered. What they considered is more specific, a
detector timing mismatch. Even if there is no detector timing
mismatch, therefore, detectors must be inspected for all possible
ways in order to get security. Giving security proofs for the BB 84
QKD system with no-clicking by coherently dealing with all
post-selection effects including those in Refs. \cite{Got04} and
\cite{Cur04,Mak06,Qi07} is an open problem.

In conclusion, we discussed a `no-clicking' event that is harmful
for Alice and Bob in the BB 84 QKD. A no-clicking event can give
rise to a post-selection effect. Specifically, we described an
attack, which we term the quantum Trojan-dark-pony attack, where
no-clicking events are utilized in the same way as double-clicking
events are utilized in the quantum Trojan-pony attack. We discussed
how to deal with the no-clicking events: The formalism of
positive-operator-valued-measurement (POM) also applies to a
no-clicking event, as is known \cite{Cur04}.  The problem of
characterizing a no-clicking event reduces to that of identifying a
positive-operator corresponding to a no-clicking event. We discussed
how to estimate the security of the protocol, against the quantum
Trojan-dark-pony attack with a positive-operator by using a formula
involving the fraction $\Delta$ of adversarial removals of events
given in Ref. \cite{Got04}.

\acknowledgments

This work was supported by the Korea Research Foundation grant
funded by the Korean Government (MOEHRD) (KRF-2005-003-C00047) and
by the Korea Science and Engineering Foundation
(R01-2006-000-10354-0).

We would like to thank Eric Corndorf, Hoi-Kwong Lo, Ranjith Nair,
Xiang-Bin Wang, and Horace Yuen for interesting discussions.

\end{document}